\documentclass[onefignum,onetabnum,a4paper]{siamonline190516}
\pdfoutput=1
\usepackage[utf8]{inputenc}
\usepackage[T1]{fontenc}
\usepackage[english]{babel}
\usepackage{url}
\usepackage{graphicx}
\usepackage{tikz}
\usepackage{colortbl}
\usepackage{xcolor}
\definecolor{myBlue}{rgb}{0.1,0.1,0.5}
\definecolor{RedViolet}{rgb}{0.6,0,0.3}
\usepackage{subcaption}
\usepackage{xargs}
\usepackage{amsmath,amsfonts,amssymb,bm,mathptmx,dsfont}
\usepackage[sort]{cite}

\usepackage[hmargin=29mm,vmargin=37mm]{geometry}

\usepackage{enumitem}
\setlist[enumerate]{leftmargin=.5in}
\setlist[itemize]{leftmargin=.5in}

\newsiamremark{remark}{Remark}
\newsiamremark{example}{Example}

\def\PaperShortTitle{Parameter selection in Gaussian process interpolation}
\def\PaperSubtitle{an empirical study of selection criteria}

\headers{\PaperShortTitle}{S.~J. Petit, J. Bect, P. Feliot and E. Vazquez}

\title{\PaperShortTitle:\\\PaperSubtitle%
  \thanks{Accepted for publication in the SIAM/ASA Journal of Uncertainty Quantification}}

\author{%
  Sébastien J. Petit\footnotemark[3] \thanks{%
    Safran Aircraft Engines, Moissy-Cramayel, France.}
                      \and
  Julien Bect\thanks{%
        Université Paris-Saclay, CNRS, CentraleSupélec,
       Laboratoire des signaux et systèmes,
       91190, Gif-sur-Yvette, France
       (\email{firstname.lastname@centralesupelec.fr}).}
                      \and
  Paul Feliot\footnotemark[2]
                      \and
  Emmanuel Vazquez\footnotemark[3]}

\def\PaperKW{%
  Gaussian processes interpolation, %
  Model choice; %
  Parameter selection; %
  Scoring rules; %
  Likelihood; %
  Regularity}

\def\PaperAuthors{%
  Sébastien J. Petit, %
  Julien Bect, %
  Paul Feliot, %
  Emmanuel Vazquez}

\hypersetup{%
  pdftitle={\PaperShortTitle: \PaperSubtitle},
  pdfauthor={\PaperAuthors},
  pdfkeywords={\PaperKW}}

\newcommand{\setGrphPATH}[1]{\def\grphPATH{#1} \graphicspath{{\grphPATH}}}

\setGrphPATH{./figures/}

\renewcommand \mathbb {\mathds}    
\renewcommand \hat    {\widehat}   

\newcommand{\abs}[1]{\lvert#1\rvert}
\newcommand{\ns}[1]{\lVert#1\rVert}
\newcommand{\mb}[1]{\mathrm{\bf #1}}

\DeclareMathOperator{\cov}{cov}
\DeclareMathOperator{\diag}{diag}

\newcommand \du   {\mathrm{d} u}
\newcommand \EE   {\mathsf{E}}     
\newcommand \Kcal {\mathcal{K}}    
\newcommand \Ncal {\mathcal{N}}    
\newcommand \Pcal {\mathcal{P}}    
\newcommand \RR   {\mathbb{R}}     
\newcommand \one  {\mathds{1}}
\newcommand \oneb {\mathbf{1}}
\newcommand \tr   {^{\mathsf{T}}}  
\newcommand \XX   {\mathbb{X}}     
\newcommand \ZZn  {{\mb{Z}_n}}
\newcommand \zzn  {\mb{z}_n}
\newcommand \zznc {\mb{z}_n^\circ}

\newcommand \SPE  {{\mathrm{SPE}}}
\newcommand \NLPD {{\mathrm{NLPD}}}
\newcommand \CRPS {{\mathrm{CRPS}}}
\newcommand \NLL  {{\mathrm{NLL}}}
\newcommand \PL   {{\mathrm{PL}}}

\newcommand \JNLL {J^{\NLL}}
\newcommand \JPL  {J^{\PL}}

\newcommand \SRule [2] {S \left(#1, #2 \right)}
\newcommand \SSPE  [2] {S^\SPE \left(#1, #2 \right)}
\newcommand \SNLPD [2] {S^\NLPD \left(#1, #2 \right)}
\newcommand \SCRPS [2] {S^\CRPS \left(#1, #2 \right)}
\newcommand \SIS   [2] {S_{1 - \alpha}^{\mathrm{IS}} \left(#1, #2 \right)}

\begin{document} \maketitle

\begin{abstract}
  This article revisits the fundamental problem of parameter selection
  for Gaussian process interpolation. By
  choosing the mean and the covariance functions of a Gaussian process within
  parametric families, the user obtains a family of Bayesian procedures
  to perform predictions about the unknown function, and must choose a
  member of the family that will hopefully provide good predictive
  performances. We base our study on the general concept of scoring
  rules, which provides an effective
  framework for building leave-one-out selection and validation
  criteria, and a notion of extended likelihood criteria based on an
  idea proposed by Fasshauer and co-authors in~2009, which makes it possible to
  recover standard selection criteria such as, for instance, the
  generalized cross-validation criterion. Under this setting, we
  empirically show on several test problems of the literature that the
  choice of an appropriate family of models is often more important than the
  choice of a particular selection criterion (e.g., the likelihood versus
  a leave-one-out selection criterion). Moreover, our numerical results
  show  that the regularity parameter of a Matérn
  covariance can be selected effectively by most selection criteria.
\end{abstract}




\section{Introduction}
\label{sec:intro}

Regression and interpolation with Gaussian processes, or kriging, is a
popular statistical tool for non-parametric function estimation, %
originating from geostatistics and time series analysis, and later
adopted in many other areas such as machine learning and the design
and analysis of computer experiments %
(see, e.g., \cite{%
  stein1999:_interpolation_of_spatial_data,%
  santner03:_desig_analy_comput_exper,%
  rasmussen06:_gauss_proces_machin_learn} and references therein). %
It is widely used for constructing fast approximations of
time-consuming computer models, with applications to %
calibration and validation %
\cite{kennedy:2001:calibration, bayarri:2007:validation}, %
engineering design %
\cite{jones:1998:ego, forrester2008:_engineering}, %
Bayesian inference %
\cite{calderhead:2009:accelerating, wilkinson:2014}, %
and the optimization of machine learning algorithms %
\cite{bergstra:2011}---to name but a few.

A Gaussian process (GP) prior is characterized by its mean and
covariance functions. %
They are usually chosen within parametric families (for instance,
constant or linear mean functions, and Matérn covariance functions), %
which transfers the problem of choosing the mean and covariance
functions to that of selecting parameters. %
The selection is most often carried out by optimization of a criterion
that measures the goodness of fit of the predictive distributions, %
and a variety of such criteria---the likelihood function, the
leave-one-out (LOO) squared-prediction-error criterion (hereafter
denoted by LOO-SPE), and others---is available from the literature. %
The search for arguments to guide practitioners in the choice of one
particular selection criterion is the main motivation of this study.

\begin{remark}
  In a fully Bayesian approach, one does not select a particular value
  for the parameters, but instead averages the predictions over a
  \emph{posterior} distribution on the parameters
  \cite{handcock93:bayesian,
    santner03:_desig_analy_comput_exper,banerjee2003}. %
  This approach may be preferred for more robust predictions~(see,
  e.g., \cite{benassi11:robust}), but comes at a higher computational
  cost. %
  For this reason, the present article will set aside the fully
  Bayesian approach and concentrate on the plugin approach, where only
  one parameter value is chosen to carry out predictions.
\end{remark}

The two most popular methods for parameter selection are maximum
likelihood (ML) and cross-validation (CV) based on LOO criteria, %
which were introduced in the field of computer experiments by the
seminal work of \cite{currin1988:_bayesian}.  Since then, despite a
fairly large number of publications dealing with ML and CV techniques
for the selection of a GP model, the literature has remained in our
view quite sparse about the relative merits of these methods, from
both theoretical and empirical perspectives.

For instance, in the framework of interpolation and infill
asymptotics, where observations accumulate in some bounded domain,
\cite{ying91:_asymp_proper_maxim_likel_estim} and
\cite{zhang04:_incon_estim_asymp_equal_inter} show that some
combinations of the parameters of a Mat\'ern covariance function can
be estimated consistently by ML. %
Again in the case of infill asymptotics, with the additional
assumption of a one-dimensional GP with an exponential covariance
function, \cite{bachoc2016:_cv_estimation_fixed_domain_asympts} show
that estimation by LOO is also consistent. %
(Similar results exist in the expanding asymptotic framework, where
observations extend over an ever-increasing horizon.)

The practical consequences of the aforementioned results are somewhat
limited in our view because practitioners are primarily interested in
the quality of the predictions. %
Knowing that the parameters of a GP can be estimated consistently is
intellectually reassuring, but may be considered of secondary
importance. These results are indeed about the statistical model
itself but they say little about the prediction properties of the
model. %
Besides, there does not exist at present, to our knowledge, some
theoretically-based evaluation of the relative performances of ML and
CV selection criteria under infill asymptotics.

Turning now to empirical comparisons of selection criteria for GP
interpolation, the first attempt in the domain of computer experiments
can be traced back to the work of \cite{currin1988:_bayesian,
  currin1991:_bayesian}. %
The authors introduce CV and ML---which can be seen as a special kind
of cross-validation---and present some simple experiments using
tensorized covariance functions, from which they conclude that, ``Of
the various kinds of cross-validation [they] have tried, maximum
likelihood seems the most reliable''. %
Additional experiments have been conducted by several authors, but no
consensus emerges from these studies: \cite{martin2003:_kriging,
  bachoc:2013:cv-ml-misspec, sunda:2001:predictive} conclude in favor
of CV whereas
\cite{martin2005:_kriging_models_deterministic_computer_models}
advocate~ML.  The reference textbook of Santner et al. \cite[Section
3.4.2]{santner03:_desig_analy_comput_exper} echoes the conclusions of
\cite{bachoc:2013:cv-ml-misspec}.

These studies are limited to a rather small number of test functions
and covariance functions, which may explain the discrepancy in the
conclusions of those experiments. %
In particular, only~\cite{bachoc:2013:cv-ml-misspec} considers the
popular and versatile Matérn covariance functions. %
Moreover, most studies focus only on the accuracy of the posterior
mean---only \cite{sunda:2001:predictive} and
\cite{bachoc:2013:cv-ml-misspec} provide results accounting for the
quality of the posterior variance---whereas the full posterior
predictive distribution is used in most GP-based methods~(see, e.g.,
\cite{jones:1998:ego, chevalier2014fast}).

This article presents two main contributions. %
First, we improve upon the results of the literature by providing an
empirical ranking of selection criteria for GP interpolation,
according to several metrics measuring the quality of posterior
predictive distributions on a large set of test functions from the
domain of computer experiments. %
To this end, we base our study on the general concept of scoring rules
\cite{gneiting07:_stric_proper_scorin_rules_predic_estim,
  zhang10:_krigin_cross_valid}, which provides an effective framework
for building selection and validation criteria.  We also introduce a
notion of extended likelihood criteria, borrowing an idea from
Fasshauer and co-authors
\cite{fasshauer:2009:optimal-scaling,fasshauer:2011:pd-kernels} in the
literature of radial basis functions.

Second, we provide empirical evidence that the choice of an
appropriate family of models is often more important%
---and sometimes much more important, especially when the size of the
design increases---%
than the choice of a particular selection criterion (e.g., likelihood
versus LOO-SPE). %
More specifically, in the case of the Matérn family, this leads us to
assess, and ultimately recommend, the automatic selection of a
suitable value of the regularity parameter, against the common
practice of choosing beforehand an arbitrary value of this parameter.

These recommendations are complementary to those of
\cite{stein1999:_interpolation_of_spatial_data}, where a particular
emphasis is placed on choosing the regularity of the covariance
function, and \cite{chen_et_al2016:_ce_assess}, where the influence of
the design and the influence of the choice of the mean and covariance
functions are studied empirically and ranked (in the case where the
parameters are selected by maximum likelihood).

The article is organized as follows. Section~\ref{sec:framework}
briefly recalls the general framework of GP regression and
interpolation.  Section~\ref{sec:meth} reviews selection criteria for
GP model parameters. %
After recalling the general notion of scoring rules, we present a
broad variety of selection criteria from the literature.
Section~\ref{sec:numer-exper} presents experimental results on the
relative performances of these criteria. %
Section~\ref{sec:conc} presents our conclusions and perspectives.

\section{General framework}
\label{sec:framework}

Let us consider the general GP approach for a scalar-valued
deterministic computer code with input space $\XX\subseteq \RR^{d}$. %
The output of the computer code~$z:\XX \to \RR$ is modeled by a random
function $\left( Z(x) \right)_{x \in \XX}$, which, from a Bayesian
perspective, represents prior knowledge about~$z$.  If we assume that
$Z(\cdot)$ is observed on a design $\XX_n = \{ x_1, \ldots, x_n \}$ of
size $n$, the data corresponds to a sample~$\zzn \in \RR^n$ of the
random vector $\ZZn = (Z(x_1), \ldots, Z(x_n))\tr$.

The GP assumption makes it possible to easily derive posterior
distributions. More precisely, it is assumed that $Z(\cdot)$ is a
Gaussian process, with (prior) mean
$\EE(Z(x)) = \sum_{l=1}^{L}\beta_l \phi_l(x)$, where the
$\beta_1,\ldots,\beta_{L}$ are unknown regression parameters and
$\phi_1,\ldots,\phi_l$ are known regression functions, and with
(prior) covariance
$\cov \left( Z(x), Z(y) \right) = k_{\vartheta}(x, y)$, where
$\vartheta \in\varTheta\subseteq\RR^{q}$ is a vector
of~\emph{parameters}.
Throughout the article, the covariance matrix of~$\ZZn$ will be
denoted by $\mb{K}_{\vartheta}$. As is often the case in applications,
we assume that the prior mean of $Z(\cdot)$ is constant (hence,
$L = 1$ with $\phi_1 = 1$). Note also that is in accordance to the
recommendation in \cite{chen_et_al2016:_ce_assess}.

One of the most popular covariance functions for GP regression is the
anisotropic stationary Matérn covariance function
\cite{matern1986:_spatial_variations} popularized by
\cite{stein1999:_interpolation_of_spatial_data}:
\begin{equation}
  \label{eq:matern-cov}
  k_{\vartheta}(x, y) = \sigma^2\, \frac{2^{1 - \nu}}{\Gamma(\nu)}
  \left( \sqrt{2 \nu} h \right)^{\nu} \Kcal_{\nu} \left( \sqrt{2 \nu} h \right),\quad
  h = \biggl(\sum_{j = 1}^d \frac{(x_j - y_j)^2}{\rho_j^2}\biggr)^{1/2},
\end{equation}
where $\Gamma$ is the Gamma function, $\Kcal_{\nu}$ is the
modified Bessel function of the second kind, and $\vartheta$ denotes
the vector of parameters
$\vartheta = (\sigma^2,\, \rho_1,\,\ldots,\, \rho_d,\, \nu) \in
\varTheta = \left( 0, \infty \right)^{d+2}$. The parameter $\sigma^2$
is the variance of~$Z(\cdot)$, the $\rho_i$s are range parameters
which characterize the typical correlation length on each dimension,
and $\nu$ is a regularity parameter, whose value controls the
mean-square differentiability of~$Z(\cdot)$. Recall (see
Table~\ref{table:pop_mat}) that the Matérn covariance function with
$\nu=1/2$ corresponds to the so-called exponential covariance
function, and the limiting case $\nu\to\infty$ can be seen as the
squared exponential (also called Gaussian) covariance function.

Because $\Kcal_{\nu}$ has a closed-form expression when
$\nu - \frac{1}{2}$ is an integer, and is more expensive to evaluate
numerically in other cases, most implementations choose to restrict
$\nu$ to half-integer values. Moreover, a widespread practice (in
applications and research papers) consists in selecting a particular
value for $\nu$ (e.g., $\nu=1/2$, $\nu=3/2$\ldots\ or the limiting
case $\nu\to\infty$), once and for all.

\begin{table}
  \caption{Popular Matérn subfamilies}
  \label{table:pop_mat}
  \centering
  \begin{tabular}{| l | c  | c | c | c |}
    \hline
    & \footnotesize $\nu = \frac{1}{2}$
    & \footnotesize $\nu = \frac{3}{2}$
    & \footnotesize $\nu = \frac{5}{2}$
    & \footnotesize $\nu = +\infty$ \\
    \hline
    \footnotesize a.k.a.
    & \footnotesize exponential & &
    & \footnotesize squared exponential\\
    \hline
    \footnotesize   $k_{\vartheta}(x, y)$
    & \footnotesize $\sigma^2 e^{-h}$
    & \footnotesize $\sigma^2 (1 + \sqrt{3}h) e^{-\sqrt{3}h}$
    & \footnotesize $\sigma^2 (1 + \sqrt{5}h + \frac{5 h^2}{3}) e^{-\sqrt{5}h}$
    & \footnotesize $\sigma^2 e^{-\frac{h^2}{2}}$  \\
    \hline
  \end{tabular}
\end{table}

Since the family of Mat\'ern covariance functions is widely used in
practice, we focus exclusively on this family in this work.

Once a GP model has been chosen, the framework of Gaussian process
interpolation allows one to build predictive distributions. More
precisely, denoting by
$\theta = \left(\beta, \vartheta\right) \in \RR \times \left(0, \, +
  \infty\right)^{d+2}$ the vector of the parameters of the model, the
usual predictive distribution for an unobserved $Z(x)$ at
$x \in \RR^{d}$ is the Gaussian posterior distribution
$\Ncal(\mu_{\theta}(x), \sigma_{\theta}^2(x))$ , where
\begin{equation}
  \label{eq:posterior-mean-var}
  \left\{\begin{array}{lll}
    \mu_{\theta}(x) &=&  \beta  + \mb{k}_{\vartheta}(x)\tr \mb{K}_{\vartheta}^{-1} \zznc,\\
    \sigma_{\theta}^2(x) &=& k_{\vartheta}(x, x) -
                             \mb{k}_{\vartheta}(x)\tr \mb{K}_{\vartheta}^{-1} \mb{k}_{\vartheta}(x)
  \end{array}\right.
\end{equation}
with
$\mb{k}_{\vartheta}(x) = (k_{\vartheta}(x, x_1),\,\ldots,\,
k_{\vartheta}(x,x_n))\tr$ and $\zznc = \zzn - \beta \oneb_n$, and
where $\oneb_n = \left(1, \dots, 1\right)\tr \in \RR^n$.

Using this framework, the user obtains a family of Bayesian
procedures, indexed by $\theta$, to perform predictions about the
unknown computer code at hand, and must choose a member of the family
that will hopefully provide good predictive performances.

\section{Selection of a GP model from a parameterized family}
\label{sec:meth}

\subsection{Scoring rules}
\label{sec:selection}

Goodness-of-fit criteria for probabilistic predictions have been
studied in the literature under the name of scoring rules by
\cite{gneiting07:_stric_proper_scorin_rules_predic_estim}.  A
(negatively oriented) scoring rule is a function
$\SRule{\,\cdot\,}{z} : \Pcal \to \RR\cup\left\{-\infty,
  +\infty\right\}$, acting on a class~$\Pcal$ of probability
distributions on~$\RR$, such that $\SRule{P}{z}$ assigns a loss for
choosing a predictive distribution $P\in\Pcal$, while observing
$z\in\RR$.  Scoring rules make it possible to quantify the quality of
probabilistic predictions.

\begin{example}[\textbf{squared prediction error}]
  Denoting by $\mu$ the mean of a predictive distribution $P$, the
  squared prediction error
  \begin{equation}
    \label{eq:SSPE}
    \SSPE{P}{z} = (z - \mu)^2
  \end{equation}
  accounts for the deviation of $z$ from $\mu$. Note that $S^\SPE$
  ignores subsequent moments and therefore predictive uncertainties.
\end{example}

\begin{example}[\textbf{negative log predictive density}]
  Denoting by $p$ the probability density of $P$,
  \begin{equation}
    \label{eq:SNLPD}
    \SNLPD{P}{z} = -\log(p(z))\,
  \end{equation}
  tells how likely $z$ is according to $P$. %
  It can be proved~\cite{bernardo:1979:utility} that any (proper)
  scoring rule that only depends on~$p(z)$ can be reduced
  to~$S^\NLPD$.
\end{example}

\begin{example}[\textbf{continuous ranked probability score}]
  Let $U$ and $U^{\prime}$ be two independent random variables with
  distribution $P$. The CRPS quantifies the deviation of $U$ from $z$:
  \begin{equation}
    \label{eq:SCRPS}
    \SCRPS{P}{z}
    = \EE(|U - z|) - \frac{1}{2} \EE(|U - U^{\prime}|)\,.
  \end{equation}
  Since
  $\SCRPS{P}{z} = \int \left(P(U \leq u)- \one_{z \leq u} \right)^2
  \du$, the CRPS can also be seen as a (squared) distance between
  the empirical cumulative distribution $u\mapsto \one_{z \leq u}$ and
  the cumulative distribution of $P$ \cite{szekely_2005:_test}.

  Note that if absolute values in~\eqref{eq:SCRPS} are replaced by
  squared values, then $S^\SPE$ is recovered. The CRPS can also be
  extended to the so-called energy and kernel scores
  \cite{gneiting07:_stric_proper_scorin_rules_predic_estim} by
  observing that $(x, y) \mapsto \abs{x - y}$ is a conditionally
  negative kernel.
\end{example}

\begin{example}[\textbf{interval score}]
  The interval scoring rule at level $1 - \alpha$ is defined, for
  $\alpha \in \left( 0, 1 \right)$, by
  \begin{equation}
    \label{eq:Sinterv}
    S_{1 - \alpha}^{\mathrm{IS}}(P;\, z) =
    (u - l) + \frac{2}{\alpha}(l - z) \one_{z \leq l}
    + \frac{2}{\alpha}(z - u) \one_{z > u}
  \end{equation}
  where $l$ and $u$ are the $\alpha/2$ and $1- \alpha/2$ quantiles of
  $P$.  The first term penalizes large intervals, while the second and
  third terms penalize intervals not containing~$z$.
\end{example}

When the predictive distribution $P$ is Gaussian, which is the case
when $P$ is the posterior distribution of a GP $Z$ at a given point,
the aforementioned scoring rules all have closed-form
expressions. More precisely, for $P = \Ncal(\mu, \sigma^2)$, we simply
have $\SSPE{P}{z} = (z-\mu)^2$
and~$\SNLPD{P}{z} = \frac{1}{2} \log 2\pi\sigma^2 + \frac{1}{2}
(z-\mu)^{2}/\sigma^2$.  $S_{1 - \alpha}^{\mathrm{IS}}$ can be obtained
by taking the standard expressions of the $\alpha/2$ and $1- \alpha/2$
quantiles of $P$, and it can be shown that
\begin{equation*}
  \SCRPS{P}{z}=
  \sigma \Bigl( \frac{z - \mu}{\sigma} \Bigl(2 \Phi \bigl( \frac{z - \mu}{\sigma} \bigr) - 1 \Bigr)
  + 2 \phi \bigl( \frac{z - \mu}{\sigma} \bigr) - \frac{1}{\sqrt{\pi}} \Bigr),
\end{equation*}
where $\phi$ and $\Phi$ stand respectively for the probability density
function and the cumulative distribution function of the standard
Gaussian distribution.

Note that all aforementioned scoring rules penalize large values of
$\abs{z - \mu}$.  When $\abs{z - \mu} \ll 1$ different scoring rules
yield different penalties, as reported in
Table~\ref{table:scoring_rules_behavior}.

\begin{table}  \newcommand\PFFFF{\vphantom{\Big(}}
  \small
  \centering
  \caption{
    Scoring rules behavior as $\abs{\mu - z} \ll 1$.
  }
  \label{table:scoring_rules_behavior}
  \begin{tabular}{| l | c  | c |  c |}
    \hline
    \PFFFF & $\sigma \ll \abs{\mu - z}$
    & $\sigma \simeq \abs{\mu - z}$
    & $\sigma \gg \abs{\mu - z}$ \\
    \hline
    \PFFFF $\SSPE{P}{z}$ & $0$ & $0$ & $0$ \\
    \PFFFF $\SNLPD{P}{z}$ & $+\infty$ & $-\infty$ & $\log \left( \sqrt{2 \pi} \sigma \right)$ \\
    \PFFFF $\SCRPS{P}{z}$ & $0$ & $0$ & $\propto \sigma$ \\
    \PFFFF $\SIS{P}{z}$ &  $0$ & $0$ & $\propto \sigma$\\
    \hline
  \end{tabular}
\end{table}

\subsection{Selection criteria}
\label{sec:calbr}

\subsubsection{Leave-one-out selection criteria}
\label{sec:cross-valid-leave}

Scoring rules make it possible to build criteria for choosing the
parameters of a GP. More precisely, to select $\theta$ based on a
sample $\zzn = \left( z_1,\, \ldots,\, z_n \right)\tr$, one can
minimize the mean score
\begin{equation}
  \label{eq:loo_gof}
  J_n^S(\theta) = \frac{1}{n}\sum_{i=1}^n S(P_{\theta, -i},\, z_i),
\end{equation}
where $S$ is a scoring rule and $P_{\theta, -i}$ denotes the
distribution of~$Z(x_i)$ conditional on~$Z(x_j) = z_j$, $j \neq i$.

Selection criteria written as~\eqref{eq:loo_gof} correspond to the
well-established leave-one-out (LOO) method, which has been proposed
in the domain of computer experiments by \cite{currin1988:_bayesian},
and is now used in many publications (see, e.g.,
\cite{rasmussen06:_gauss_proces_machin_learn}, and also
\cite{zhang10:_krigin_cross_valid}, who formally adopt the point of
view of the scoring rules, but for model validation instead of
parameter selection).

\paragraph{Efficient computation of predictive distributions} %
Leave-one-out predictive densities can be computed using fast
algebraic formulas \cite{craven1978:_smoothing_noisy_data,
  dubrule1983:_cv_kriging}. More precisely, the predictive
distribution $P_{\theta,-i}$ is a normal distribution
$\Ncal(\mu_{\theta,-i}, \sigma_{\theta,-i}^2)$ with
\begin{equation}
  \label{eq:LOO_formulas}
  \mu_{\theta, -i} = z_i - \frac{ \left( \mb{K}_{\vartheta}^{-1} \zznc \right)_i}{( \mb{K}_{\vartheta}^{-1} )_{i, i}} \quad \mathrm{and}
  \quad \sigma_{\theta, -i}^2 = \frac{1}{( \mb{K}_{\vartheta}^{-1} )_{i, i}}.
\end{equation}

\begin{remark}
  Equation~\eqref{eq:LOO_formulas} provides a numerically convenient
  expression for~$\mu_{\theta, -i}$, which apparently
  involves~$z_i$. %
  However, $z_i$ appears twice in the expression, and the two
  occurences cancel out when the matrix product is written explicitly.
\end{remark}

Furthermore, \cite{petit20:_towar_gauss} show that, using reverse-mode
differentiation, it is possible to compute mean scores $J_n$ and their
gradients with a $O(n^3 + dn^2)$ computational cost, which is the same
computational complexity as for computing the likelihood function and
its
gradient~(see,~e.g.,~\cite{rasmussen06:_gauss_proces_machin_learn}).

\paragraph{The particular case of LOO-SPE} %
The LOO selection criterion
\begin{equation}
  \label{eq:LOO_SPE}
  J_n^\SPE (\theta) = \frac{1}{n} \sum_{i = 1}^n (\mu_{\theta, -i} - {z_i})^2\,,
\end{equation}
based on the scoring rule~\eqref{eq:SSPE} will be referred to as
LOO-SPE.  This criterion, also called prediction sum of squares
(PRESS) \cite{allen1974:_press, wahba1990:_splines_models} or LOO
squared bias \cite{currin1988:_bayesian}, is well known in statistics
and machine learning, and has been advocated by some authors
\cite{sunda:2001:predictive, martin2003:_kriging,
  bachoc:2013:cv-ml-misspec, bachoc:2018:asympt-misspec,
  santner03:_desig_analy_comput_exper} to address the case of
``misspecified'' covariance functions.

However, note that $\sigma^2$ cannot be selected using
$J_n^{\rm SPE}$. When $J_n^\SPE$ is used, $\sigma^2$ is generally
chosen~(see, e.g., \cite{currin1988:_bayesian,
  bachoc:2013:cv-ml-misspec} and Remark~\ref{rmq:sigma_square}) to
satisfy
\begin{equation}
  \label{eq:cressie}
  \frac{1}{n} \sum_{i=1}^n \frac{(z_i -\mu_{\theta,
      -i})^2}{\sigma_{\theta, -i}^2 } = 1\,,
\end{equation}
which will be referred to as Cressie's rule for $\sigma^2$, in
reference to the claim by \cite{cressie93:_statis_spatial_data} that
\eqref{eq:cressie} should hold approximately for a good GP model.

\paragraph{Other scoring rules for LOO} %
The selection criteria using the NLPD scoring rule~\eqref{eq:SNLPD}
and the CRPS scoring rule~\eqref{eq:SCRPS} will be referred to as the
LOO-NLPD and LOO-CRPS criteria, respectively. %
The LOO-NLPD criterion has been called preditive deficiency
in~\cite{currin1988:_bayesian}, and Geisser's surrogate Predictive
Probability (GPP) in~\cite{sunda:2001:predictive}. %
The LOO-CRPS criterion has been considered
in~\cite{zhang10:_krigin_cross_valid} as a criterion for model
validation (see also~\cite{demay:hal-03207216} for an application to
model selection), and more recently \cite{petit20:_towar_gauss,
  petit:mascot-2020} as a possible criterion for parameter selection
as well.

\begin{remark} \label{rmq:sigma_square} %
  Note that Cressie's rule~\eqref{eq:cressie} can be derived by
  minimizing the LOO-NLPD criterion with respect to $\sigma^2$.
\end{remark}

\begin{remark}
  In order to limit the number of selection criteria under study, the
  interval scoring rule is only used for validation in this work.
\end{remark}

\subsubsection{Maximum likelihood and generalizations}
\label{sec:maximum-likelihood}

We can safely say that the most popular method for selecting $\theta$
from data is maximum likelihood estimation---and related techniques,
such as restricted maximum likelihood estimation.  The ML estimator is
obtained by maximizing the likelihood function or, equivalently, by
minimizing the negative log-likelihood (NLL) selection
criterion. Denoting by $p_{\theta}(\zzn)$ the joint density of~$\ZZn$,
the NLL selection criterion may be written as
\begin{equation}
  \label{eq:NLL}
  \JNLL_n(\theta) = -\log(p_{\theta}(\zzn)) =
  \frac{1}{2}\left( n\log(2\pi) + \log\det \mb{K}_{\vartheta} + \left( \zznc \right)\tr  \mb{K}_{\vartheta}^{-1}
    \zznc \right)\,.
\end{equation}

As pointed out by \cite{currin1988:_bayesian}, the NLL criterion is
closely related to the LOO-NLPD criterion, through the identity %
\begin{equation*}
  \JNLL_n(\theta) =
  - \log( p_{\theta}(z_1))
  - \sum_{i=2}^n \log ( p_{\theta}(z_i \mid z_1, \ldots, z_{i-1})
  )),
\end{equation*}
which makes appear the densities of the $Z(x_i)$s conditional on
$Z(x_j)=z_j$, $j<i$ (whereas the LOO-NLPD criterion involves the
densities of the $Z(x_i)$s conditional on $Z(x_j)=z_j$, $j \neq i$).

One can minimize~\eqref{eq:NLL} in closed-form with respect to
$\sigma^2$, given other parameters. %
Writing $\mb{K}_{\vartheta} = \sigma^2\, \mb{R}_{\vartheta}$ and
canceling
${\partial \JNLL_n(\theta)}/{\partial \sigma^{2}} = \bigl(n \sigma^2-
\left( \zznc \right)\tr \mb{R}_{\vartheta}^{-1} \zznc \bigr) /
(2\sigma^2)$ yields
\begin{equation}\label{eq:ml_rule}
  \sigma^2_{\NLL} = \frac{1}{n} \left( \zznc \right)\tr  \mb{R}_{\vartheta}^{-1} \zznc\,,
\end{equation}
which will be referred to as the profiling rule for $\sigma^2$.

Injecting~\eqref{eq:ml_rule} into~\eqref{eq:NLL} yields a
\emph{profile likelihood} selection criterion, that can be written as
\begin{equation}\label{eq:prof_nll}
  \JPL_n(\theta) =   \log \sigma_{\NLL}^2 + \frac{1}{n}\log\det
  \mb{R}_{\vartheta} = \log \Bigl(\frac{1}{n} \left( \zznc \right)\tr
  \mb{R}_{\vartheta}^{-1} \zznc\Bigr)  + \frac{1}{n}\log\det \mb{R}_{\vartheta} \,.
\end{equation}

Following Fasshauer and co-authors
\cite{fasshauer:2009:optimal-scaling,fasshauer:2011:pd-kernels}, we
consider now a family of selection criteria that
extends~\eqref{eq:NLL}.  Using the factorization
$\mb{R}_{\vartheta} = \mb{Q} {\Lambda} \mb{Q}\tr$, where
$\mb{Q} = (\mb{q}_1,\, \ldots,\, \mb{q}_n)$ is an orthogonal matrix of
(orthonormal) eigenvectors and
${\Lambda} = \diag(\lambda_1, \cdots, \lambda_n)$, notice that
\begin{equation}
  \label{eq:NLL-fact}
  \begin{aligned}
    \exp\left(\JPL_n(\theta)\right)
    & \;=\; \frac{1}{n} \left( \zznc \right)\tr
    \mb{R}_{\vartheta}^{-1} \zznc \cdot (\det\mb{R}_{\vartheta})^{1/n}\\
    & \;\propto\; \Bigl( \sum_{i = 1}^n \left(\bm{q}_i\tr \zznc\right)^2/ \lambda_i \Bigr)
    \Bigl( \prod_{i=1}^n  \lambda_j \Bigr)^{1/n}.
  \end{aligned}
\end{equation}
This suggests a generalization of the likelihood criterion that we
shall call Fasshauer's Hölderized likelihood (HL), defined as
\begin{equation}\label{eq:crit-HL}
    J_n^{\mathrm{HL},\, p,\,q}(\theta) =
    \Bigl( \sum_{i = 1}^n (\bm{q}_i\tr \zznc)^2/ \lambda_i^p \Bigr)^{1/p}
    \Bigl( \frac{1}{n} \sum_{j = 1}^n \lambda_j^q \Bigr)^{1/q},
\end{equation}
with $q \in [-\infty, +\infty]$, and $p \in \RR \setminus \{0\}$, and
where $\sigma^2$ can be chosen using the rules~\eqref{eq:cressie}
or~\eqref{eq:ml_rule}, since $J_n^{\mathrm{HL},\, p,\,q}(\theta)$ does
not depend on $\sigma^2$. Owing to the standard property of
generalized means
$\left(\frac{1}{n}\sum_{i=1}^{n}x_i^q\right)^{\frac{1}{q}}
\stackrel{q\to 0}{\longrightarrow} \sqrt[n]{x_1\cdots x_n}$),
\eqref{eq:NLL-fact} is recovered by taking $p = 1$ and letting
$q \to 0$. Moreover, two other known selection criteria can be
obtained for particular values of $p$ and $q$, as detailed below.

\paragraph{Generalized cross-validation} Taking $p = 2$ and $q = -1$
in~\eqref{eq:crit-HL} yields the generalized cross-validation (GCV)
criterion
\begin{equation}
  J_n^{\mathrm{GCV}}(\theta) = n^{-1} \left( J_n^{\mathrm{HL},\, 2,\,-1}(\theta) \right)^2,  
\end{equation}
which was originally proposed as a rotation-invariant version of PRESS
\cite{golub1979:_gcv_ridge} for linear models.  It has also been shown
to be efficient for the selection of the smoothing parameter of
polyharmonic splines~\cite{wahba1990:_splines_models} and for the
selection of the degree of a
spline~\cite{wahba1980:_mathematical_methods_splines_cv}.

The GCV selection criterion is a weighted SPE criterion, which can
also be written as
\begin{equation}
  \label{eq:GCV}
  J_n^{\mathrm{GCV}}(\theta) = \frac{1}{n} \sum_{i=1}^n w_i^{2}(\theta)
  (z_i - \mu_{\theta, -i})^2\,, \quad w_i(\theta)
  = \frac{\tilde \sigma^{2}}{\sigma_{\theta, -i}^2}\,,
\end{equation}
with
$\tilde \sigma^2 = \bigl(\frac{1}{n} \sum_{i=1}^n
\frac{1}{\sigma_{\theta, -i}^2}\bigr)^{-1}$.  Notice that
$w_i(\theta)$ is lower when $\sigma_{\theta, -i}$ is larger, which
happens when points are either isolated or lying on the border /
envelope of $\XX_n$.  Equation~\eqref{eq:GCV} shows that, similarly to
the LOO criteria of Section~\ref{sec:cross-valid-leave}, the GCV
criterion can be computed, along with its gradient, in $O(n^3 + dn^2)$
time.  However, since $\tilde \sigma$ depends on the data, the
$J_n^{\rm GCV}$ criterion cannot be formally derived from a scoring
rule.

\paragraph{Kernel alignment} The kernel alignment (KA) selection
criterion\cite{cristianini01:_kernel_target_align}
\begin{equation}
  \label{eq:KA}
  J_n^{\mathrm{KA}}(\theta) =
  -\frac{ \left( \zznc \right)\tr \mb{K}_{\vartheta} \zznc}{\ns{\mb{K}_{\vartheta}}_{F} \ns{\zznc}^2}\,,
\end{equation}
where $\ns{\,\cdot\,}_{F}$ stands for the Frobenius matrix
norm, measures the (opposite of the) alignment of~$\zznc$ with the eigenvector of
  $\mb{K}_{\vartheta}$ corresponding to the largest eigenvalue. %
This criterion is related to~\eqref{eq:crit-HL} with $p = -1$
and $q = 2$:
\begin{equation}
  J_n^{\mathrm{KA}}(\theta) = - \frac{1}{\sqrt{n} \ns{\zznc}^2
  J_n^{\mathrm{HL},\, -1,\, 2}(\theta) }.
\end{equation}
Thus minimizing the KA criterion~\eqref{eq:KA} is equivalent to
  minimizing $J_n^{\mathrm{HL},\, -1,\, 2}$ when the mean~$\beta$ is known
  (otherwise~$\zznc$ also depends on~$\theta$). %
The KA criterion is noticeably cheaper than the
others, as it does not require to invert $\mb{K}_{\vartheta}$ and can
therefore be computed along with its gradient in $O(dn^2)$ time.

\begin{remark}
  \label{rem:KA-mean}
  The KA criterion as written above is only suitable
  for selecting covariance parameters, and cannot be used to select both the mean~$\beta$
  and the range parameters as is done in this study. Indeed, %
  the criterion is minimized when both $\beta$ and the range
  parameters~$\rho_j$ go to infinity. .
\end{remark}

\begin{remark}
  We choose to focus in this article on two well-known selection
  criteria (NLL and GCV) that can be seen as special cases
  of~\eqref{eq:crit-HL}, corresponding repectively to $(p, q)$ equal
  to $(1, 0)$, $(2, -1)$. %
  The study of new selection criteria, obtained for other values
  of~$(p, q)$, is left for future work.
\end{remark}

\subsection{Hybrid selection criteria}\label{sec:hybrid}

When considering several parameterized models---or, equivalently, when
dealing with discrete parameters, such as half-integer values for the
regularity parameter of the Matérn covariance---some authors suggest
to use one selection criterion to select the parameters in each
particular model, and a different one to select the model itself.

For instance, in~\cite{jones:1998:ego}, the authors select the
parameters of a power-exponential covariance function using the NLL
selection criterion (i.e., the ML method), and then select a suitable
transformation of the output of the simulator, in a finite list of
possible choices, using the LOO-SPE criterion. %
Similarly, the NLL selection criterion is combined
in~\cite{demay:hal-03207216} with a variety of model-validation
criteria, including LOO-CRPS and LOO-NLPD.

In Section~\ref{sec:numer-exper} we will denote by NLL/SPE the hybrid
method that selects the variance and range parameters of a Matérn
covariance function using the NLL criterion, and then minimizes the
LOO-SPE criterion to select the regularity parameter~$\nu$ in finite
list of values.

\section{Numerical experiments}
\label{sec:numer-exper}

\subsection{Methodology}
\label{sec:methodology}

We investigate the problem of parameter selection with an experimental
approach consisting of four ingredients: 1)~a~set of unknown functions
$z$ to be predicted using evaluation results on a finite design
$\XX_n = \{x_1, \ldots, x_n\}\subset\XX$; 2)~the GP regression method
that constructs predictive distributions $P_{\theta,\,x}$ of~$z$ at
given $x$s in $\XX$, indexed by parameter $\theta$; 3)~several
selection criteria $J_n$ to choose $\theta$; 4)~several criteria to
assess the quality of the $P_{\theta,\,x}$s. Details about each of
these ingredients are given below (starting from the last one).

\paragraph{Criteria to assess the quality of the $P_{\theta,\,x}$s}
A natural way to construct a criterion to assess the quality of the
$P_{\theta,\,x}$s is to choose a scoring rule $S$ and to consider the
mean score on a test set
$\XX_N^{\text{test}}=\{x_1^{\text{test}},\,\ldots,\,x_N^{\text{test}}\}\subset
\XX$ of size $N$:
\begin{equation}
  \label{eq:gof}
  R(\theta;\, S) = \frac{1}{N}\sum_{i=1}^N S\bigl(P_{\theta,\, x_i^{\text{test}}},\,
  z(x_i^{\text{test}})\bigr)\,.
\end{equation}

\paragraph{Selection criteria} We shall consider the selection
criteria $J_n$ presented in Section~\ref{sec:meth}, namely, the
LOO-SPE, LOO-NLPD, LOO-CRPS, NLL, GCV, and NLL/SPE selection
criteria. Given a function $f$ and a design $\XX_n$, each selection
criterion $J_n$ yields a parameter~$\theta_{J_n}$.

\begin{remark}
  We do not include KA in our list of selection criteria since it is
  not suitable to select the mean~$\beta$ (cf.\
  Remark~\ref{rem:KA-mean}). %
  Note that experiments in the zero-mean case \cite{petit2022:improved}
  suggest that the KA criterion performs poorly when compared to the other criteria
  considered in the article.
\end{remark}

\paragraph{Parameterized GP models} In this work, models are
implemented using a custom version of the GPy \cite{gpy2014} Python
package (see Supplementary Material, hereafter abbreviated as SM). %
We assume no observation noise, which corresponds to the interpolation
setting. %
A constant mean function and the anisotropic Matérn covariance
function~\eqref{eq:matern-cov} are used. The parameter, to be
selected, is
$\theta = (\beta, \sigma^2,\, \rho_1,\, \ldots,\, \rho_d,\, \nu)$. %
The regularity parameter~$\nu$ is either set a priori to
$\nu = \chi + 1/2$, with
$\chi \in \{ 0, 1, 2, 3, 4, d, 2d, \infty \}$, or selected
automatically to a value denoted by~$\hat\nu$.

\begin{remark}\label{rem:bounds}
  Since the covariance matrix of~$\ZZn$ can be nearly singular when
  the range parameters take large values, we define upper bounds on
  these values in order to avoid the use of \emph{nugget} or
  \emph{jitter} (see, e.g., \cite{ranjan:2011:comput-stable,
    peng:_nugget_choice}).  Details are provided in the SM.
\end{remark}

\paragraph{Test functions} %
The test functions used in the study are described in the next
section. %
They are grouped into collections, and we provide averaged values of
mean-score metrics of the form~\eqref{eq:gof} for each collection.

\subsection{Test functions}
\label{sec:test-functions}

\subsubsection{Design of a low-pass filter}
\label{sec:low-pass-filter-descr}

\begin{figure}
\begin{center}
\includegraphics[scale=0.345]{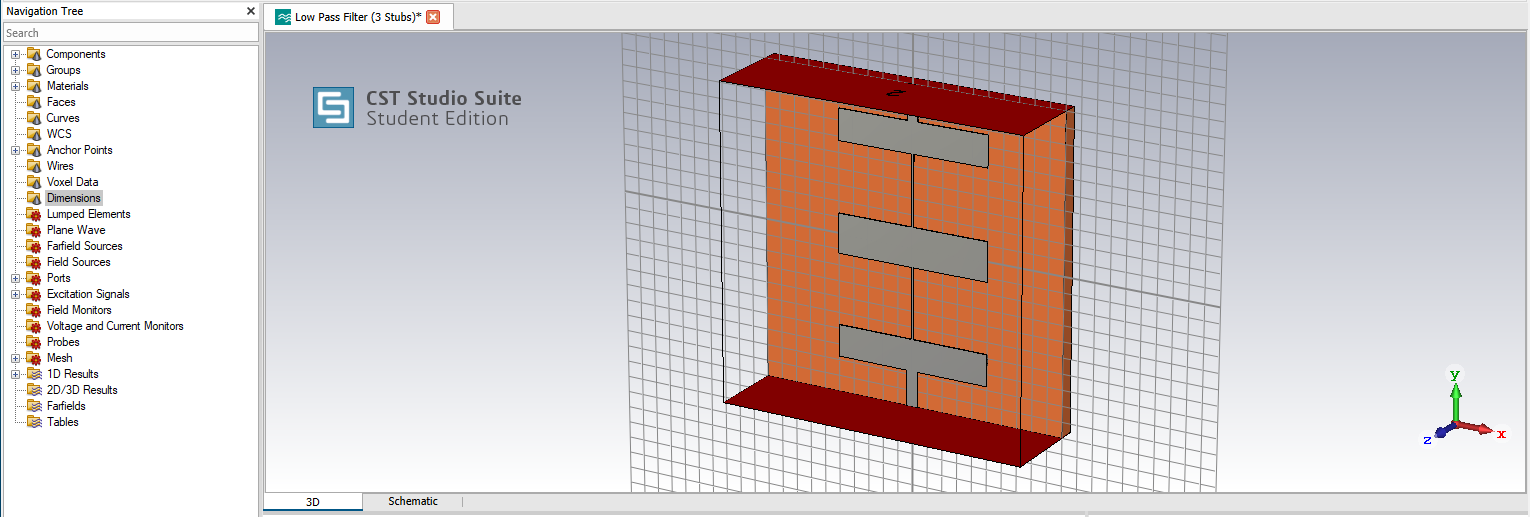}
\end{center}
\caption{A low-pass filter design problem in CST Studio
Suite\textregistered.
}
\label{fig:low-pass}
\end{figure}

Fuhrländer and Schöps \cite{fuhrlander:2020:yield} consider the
problem of computing, using a frequency-domain PDE solver, the
scattering parameters~$s_{\omega}$ of an electronic component called
stripline low-pass filter, at several values of the excitation
pulsation~$\omega$. %
The geometry of the stripline filter is illustrated in
Figure~\ref{fig:low-pass}. It is parameterized using six real valued
factors concatenated in a vector $x\in\RR^{d}$, $d=6$. The objective
is to satisfy the low-pass specifications
$|s_{2k\pi}(x)| \geq - 1\mathrm{dB}$ for $0 \leq k \leq 4$ and
$|s_{2k\pi}(x)| \leq -20\mathrm{dB}$ for $5 \leq k \leq 7$. Meeting
such requirements is a difficult and time-consuming task.

In this article, we consider the quantities $\mathrm{Re}(s_{2\pi})$,
$\mathrm{Re}(s_{6\pi})$, $\mathrm{Re}(s_{10\pi})$, and
$\mathrm{Re}(s_{14\pi})$, where $\mathrm{Re(z)}$ denotes the real part
of a complex number~$z$.
For three different design sizes $n \in \{10d, 20d, 50d\}$,
we randomly sample $M = 100$ subsets~$\XX_{n}$ of size~$n$ from a
database of $10000$~simulation results, and use the remaining
$N = 10000 - n$ points as test sets. %
The data are standardized so that the output values
have zero mean and unit variance on the test sets. %
The metric~\eqref{eq:gof} is computed and averaged on these $M$~repetitions.

\subsubsection{Other test functions}
\label{sec:test-functions-from}

We supplement the above engineering problem with a collection of test
functions from the literature.
More precisely, we consider the Goldstein-Price function
\cite{dixon1978:_global_optimization}, a one-dimensional version of
the Goldstein-Price function (see SM for details), the Mystery
function \cite{martin2003:_kriging}, the Borehole function
\cite{worley1987:_deterministic_uncertainty_analysis}, several
collections obtained from the GKLS simulator \cite{gaviano2003:_toms},
and the rotated Rosenbrock collection from the BBOB benchmark suite
\cite{hansen:2009:bbob}.

The GKLS simulator has a ``smoothness'' parameter~$k \in \{0, 1, 2\}$
controlling the presence of non-differentiabilities on some nonlinear
subspaces---the trajectories being otherwise infinitely
differentiable.
For both GKLS and Rosenbrock, two different values of the input
dimension were considered ($d = 2$ and $d = 5$).
The resulting set of twelve problems---considering that changing the
value of~$k$ or~$d$ defines a new problem---is summarized in Table
\ref{table:summary}.

For each problem, we consider three design sizes
$n \in \{10d, 20d, 50d\}$. %
For the GKLS and Rosenbrock collections, we used the collections of
test functions provided by the authors ($M = 100$ and $15$ functions,
respectively). %
They are evaluated on random space-filling designs~$\XX_{n}$ (see
below).

For the Goldstein-Price function, the one-dimensional version of the
Goldstein-Price function, the Mystery function and the Borehole
function, we evaluate each test function on~$M = 100$ random
space-filling designs~$\XX_n$.

Random space-filling designs are obtained using pseudo-maximin Latin
hypercube sampling obtained from $1000$~random draws (see, e.g.,
\cite{pronzato2012design}). %
Test sets are constructed using a Sobol' sequence
$\XX_{N}^{\text{test}}$ of size $N = 10000$ for the evaluations of the
functions. %
To make the results comparable, all functions are centered and
normalized to unit variance on the test sets.

\begin{table}
  \small
  \caption{Twelve benchmark problems}
  \centering \small \setlength{\tabcolsep}{5pt}
  \begin{tabular}{| l || c | c | c | c | c | c | c | r |}
    \hline
    Problem  & Goldstein-Price & Mystery & $\mathrm{GKLS}_{k=0}$ & $\mathrm{GKLS}_{k=1}$ & $\mathrm{GKLS}_{k=2}$ & Rosenbrock      & Borehole  \\
    \hline
    $d$         & $\{ 1, 2 \}$                & $2$       & $\{ 2, 5 \}$ & $\{ 2, 5 \}$ & $\{ 2, 5 \}$ & $\{ 2, 5 \}$    &  $8$ \\
    \hline
  \end{tabular}
  \label{table:summary}
\end{table}


\subsection{Results and findings}
\label{sec:results-and-findings}

\subsubsection{A close look at one of the problems}\label{sec:closer-look}

Tables~\ref{table:results-low-pass-SPE}
and~\ref{table:results-low-pass-IS} provide a detailed view of the
results obtained on one of the test problems---namely, the output
$\mathrm{Re}(s_{2\pi})$ with $n = 10 d = 60$ of the low-pass filter
case (see Section~\ref{sec:low-pass-filter-descr}).

\newcommand\myTabConfig{\footnotesize \def\arraystretch{0.9} \setlength\tabcolsep{4pt}}

\begin{table}[t]

  \caption{Averages (over the $M=100$ designs) of the
    $R(\theta; S^\SPE)$ values for $\mathrm{Re}(s_{2\pi})$ with
    $n = 10 d = 60$.  For comparison, the rightmost column gives the
    optimal value~$R^{\star}$ obtained by direct minimization of the
    score~\eqref{eq:gof}; see SM for more details.  Using
    $J_n^{\mathrm{NLL/SPE}}$ as selection criterion, which also
    selects~$\nu$ (see Section~\ref{sec:hybrid}), we have
    $R(\theta; S^\SPE) = 2.00 \cdot 10^{-5}$.  The gray scale
    highlights the order of magnitude of the discrepancies.  }

  \begin{center} \myTabConfig

\begin{tabular}{| l | c | c | c | c | c | c |}
\hline
Scoring rule: $S^{\SPE}$ &NLL & LOO-SPE & LOO-NLPD & LOO-CRPS & GCV & $R^{\star}$\\
\hline
$\nu = 1/2$ & \cellcolor[gray]{0.6}$2.56 \cdot 10^{-2}$ & \cellcolor[gray]{0.6}$2.78 \cdot 10^{-2}$ & \cellcolor[gray]{0.6}$2.43 \cdot 10^{-2}$ & \cellcolor[gray]{0.6}$2.43 \cdot 10^{-2}$ & \cellcolor[gray]{0.6}$2.35 \cdot 10^{-2}$ & $2.09 \cdot 10^{-2}$ \\
\hline
$\nu = 3/2$ & \cellcolor[gray]{0.7345694123399603}$7.41 \cdot 10^{-5}$ & \cellcolor[gray]{0.7018290087255415}$8.95 \cdot 10^{-5}$ & \cellcolor[gray]{0.7409175452057712}$7.15 \cdot 10^{-5}$ & \cellcolor[gray]{0.7273412537601565}$7.73 \cdot 10^{-5}$ & \cellcolor[gray]{0.7413254881509481}$7.13 \cdot 10^{-5}$ & $6.00 \cdot 10^{-5}$ \\
\hline
$\nu = 5/2$ & \cellcolor[gray]{0.9945375857302988}$1.66 \cdot 10^{-5}$ & \cellcolor[gray]{0.9375771954435116}$2.30 \cdot 10^{-5}$ & \cellcolor[gray]{1.0}$1.61 \cdot 10^{-5}$ & \cellcolor[gray]{0.9871780257143297}$1.73 \cdot 10^{-5}$ & \cellcolor[gray]{0.9699454355504595}$1.91 \cdot 10^{-5}$ & $1.02 \cdot 10^{-5}$ \\
\hline
$\nu = 7/2$ & \cellcolor[gray]{0.9660934839782296}$1.95 \cdot 10^{-5}$ & \cellcolor[gray]{0.8836645554444178}$3.14 \cdot 10^{-5}$ & \cellcolor[gray]{0.9300242166841347}$2.41 \cdot 10^{-5}$ & \cellcolor[gray]{0.9226230846519087}$2.51 \cdot 10^{-5}$ & \cellcolor[gray]{0.8766983445151698}$3.27 \cdot 10^{-5}$ & $1.07 \cdot 10^{-5}$ \\
\hline
$\nu = 9/2$ & \cellcolor[gray]{0.91807488485702}$2.58 \cdot 10^{-5}$ & \cellcolor[gray]{0.8312950782311791}$4.25 \cdot 10^{-5}$ & \cellcolor[gray]{0.8773947664182874}$3.26 \cdot 10^{-5}$ & \cellcolor[gray]{0.8657450807455579}$3.48 \cdot 10^{-5}$ & \cellcolor[gray]{0.7944232919222021}$5.25 \cdot 10^{-5}$ & $1.53 \cdot 10^{-5}$ \\
\hline
$\nu = 13/2$ & \cellcolor[gray]{0.8773138690239091}$3.26 \cdot 10^{-5}$ & \cellcolor[gray]{0.7604866998284127}$6.38 \cdot 10^{-5}$ & \cellcolor[gray]{0.8368632460094017}$4.11 \cdot 10^{-5}$ & \cellcolor[gray]{0.8356975180638393}$4.14 \cdot 10^{-5}$ & \cellcolor[gray]{0.6642503998960931}$1.11 \cdot 10^{-4}$ & $1.90 \cdot 10^{-5}$ \\
\hline
$\nu = 25/2$ & \cellcolor[gray]{0.8319276723021698}$4.23 \cdot 10^{-5}$ & \cellcolor[gray]{0.7498402944405814}$6.79 \cdot 10^{-5}$ & \cellcolor[gray]{0.7733447030659566}$5.93 \cdot 10^{-5}$ & \cellcolor[gray]{0.784109217860216}$5.57 \cdot 10^{-5}$ & \cellcolor[gray]{0.6567234372669878}$1.16 \cdot 10^{-4}$ & $2.30 \cdot 10^{-5}$ \\
\hline
$\nu = \infty$ & \cellcolor[gray]{0.8074729353947369}$4.87 \cdot 10^{-5}$ & \cellcolor[gray]{0.7332402277052694}$7.47 \cdot 10^{-5}$ & \cellcolor[gray]{0.7774849430614964}$5.79 \cdot 10^{-5}$ & \cellcolor[gray]{0.774309295183782}$5.90 \cdot 10^{-5}$ & \cellcolor[gray]{0.6993067135568072}$9.08 \cdot 10^{-5}$ & $2.56 \cdot 10^{-5}$ \\
\hline
$\nu \in \{ 1/2 , \cdots,\infty \}$ & \cellcolor[gray]{0.9913521605848714}$1.69 \cdot 10^{-5}$ & \cellcolor[gray]{0.8712244869701028}$3.37 \cdot 10^{-5}$ & \cellcolor[gray]{0.9514404840384824}$2.13 \cdot 10^{-5}$ & \cellcolor[gray]{0.9408343684741801}$2.26 \cdot 10^{-5}$ & \cellcolor[gray]{0.9100433356188329}$2.70 \cdot 10^{-5}$ & $8.99 \cdot 10^{-6}$ \\
\hline
\end{tabular}

\end{center}

\label{table:results-low-pass-SPE}
\end{table}

\begin{table}[t]

\caption{Same as Table \ref{table:results-low-pass-SPE} but
  for averages of $R(\theta; S_{0.95}^{\mathrm{IS}} )$.
  Using $J_n^{\mathrm{NLL/SPE}}$ gives $R(\theta; S_{0.95}^{\mathrm{IS}}) = 2.18 \cdot 10^{-2}$.
}

  \begin{center} \myTabConfig

\begin{tabular}{| l | c | c | c | c | c | c |}
\hline
Scoring rule: $S_{0.95}^{\mathrm{IS}}$ &NLL & LOO-SPE & LOO-NLPD & LOO-CRPS & GCV\\
\hline
$\nu = 1/2$ & \cellcolor[gray]{0.6}$1.05 \cdot 10^{0}$ & \cellcolor[gray]{0.6}$9.86 \cdot 10^{-1}$ & \cellcolor[gray]{0.6}$9.10 \cdot 10^{-1}$ & \cellcolor[gray]{0.6}$1.03 \cdot 10^{0}$ & \cellcolor[gray]{0.6}$8.91 \cdot 10^{-1}$ \\
\hline
$\nu = 3/2$ & \cellcolor[gray]{0.8468295681530961}$4.25 \cdot 10^{-2}$ & \cellcolor[gray]{0.8384194930621747}$4.46 \cdot 10^{-2}$ & \cellcolor[gray]{0.8634953489538161}$3.86 \cdot 10^{-2}$ & \cellcolor[gray]{0.8534997646037159}$4.09 \cdot 10^{-2}$ & \cellcolor[gray]{0.8653153163284162}$3.82 \cdot 10^{-2}$ \\
\hline
$\nu = 5/2$ & \cellcolor[gray]{0.9857454609483716}$1.91 \cdot 10^{-2}$ & \cellcolor[gray]{0.9697635105203714}$2.09 \cdot 10^{-2}$ & \cellcolor[gray]{1.0}$1.76 \cdot 10^{-2}$ & \cellcolor[gray]{0.9882902548099015}$1.88 \cdot 10^{-2}$ & \cellcolor[gray]{0.9887509389375426}$1.88 \cdot 10^{-2}$ \\
\hline
$\nu = 7/2$ & \cellcolor[gray]{0.9713768255866562}$2.08 \cdot 10^{-2}$ & \cellcolor[gray]{0.9388777988646089}$2.50 \cdot 10^{-2}$ & \cellcolor[gray]{0.964070720063118}$2.16 \cdot 10^{-2}$ & \cellcolor[gray]{0.9458175835093889}$2.40 \cdot 10^{-2}$ & \cellcolor[gray]{0.9384078149168065}$2.51 \cdot 10^{-2}$ \\
\hline
$\nu = 9/2$ & \cellcolor[gray]{0.9372770361107203}$2.53 \cdot 10^{-2}$ & \cellcolor[gray]{0.8995235782004029}$3.14 \cdot 10^{-2}$ & \cellcolor[gray]{0.9170648017855405}$2.84 \cdot 10^{-2}$ & \cellcolor[gray]{0.8963428669675955}$3.20 \cdot 10^{-2}$ & \cellcolor[gray]{0.8785740944574897}$3.54 \cdot 10^{-2}$ \\
\hline
$\nu = 13/2$ & \cellcolor[gray]{0.9047758560152744}$3.04 \cdot 10^{-2}$ & \cellcolor[gray]{0.8566283626494126}$4.02 \cdot 10^{-2}$ & \cellcolor[gray]{0.8781599171778018}$3.55 \cdot 10^{-2}$ & \cellcolor[gray]{0.8622014385056953}$3.89 \cdot 10^{-2}$ & \cellcolor[gray]{0.8334397568994772}$4.59 \cdot 10^{-2}$ \\
\hline
$\nu = 25/2$ & \cellcolor[gray]{0.8710988802054915}$3.70 \cdot 10^{-2}$ & \cellcolor[gray]{0.8417835299056119}$4.38 \cdot 10^{-2}$ & \cellcolor[gray]{0.8383408126102984}$4.46 \cdot 10^{-2}$ & \cellcolor[gray]{0.833429381865735}$4.59 \cdot 10^{-2}$ & \cellcolor[gray]{0.8107206518874502}$5.23 \cdot 10^{-2}$ \\
\hline
$\nu = \infty$ & \cellcolor[gray]{0.8506263778598041}$4.16 \cdot 10^{-2}$ & \cellcolor[gray]{0.8292752186647315}$4.70 \cdot 10^{-2}$ & \cellcolor[gray]{0.8327637278559498}$4.61 \cdot 10^{-2}$ & \cellcolor[gray]{0.8206530812283271}$4.94 \cdot 10^{-2}$ & \cellcolor[gray]{0.7983940388299551}$5.62 \cdot 10^{-2}$ \\
\hline
$\nu \in \{ 1/2 , \cdots,\infty \}$ & \cellcolor[gray]{0.9825865727838287}$1.95 \cdot 10^{-2}$ & \cellcolor[gray]{0.9181967109793711}$2.82 \cdot 10^{-2}$ & \cellcolor[gray]{0.9652495128641723}$2.15 \cdot 10^{-2}$ & \cellcolor[gray]{0.9440378103021174}$2.43 \cdot 10^{-2}$ & \cellcolor[gray]{0.938851799708815}$2.50 \cdot 10^{-2}$ \\
\hline
\end{tabular}
\end{center}

\label{table:results-low-pass-IS}
\end{table}

The results presented in these tables are the scores $R(\theta; S)$,
averaged over the $M = 100$ repetitions, where
$\theta$ is selected using different selection criteria (along
columns), and the regularity of the Matérn covariance varies or is
selected automatically (along rows). %
The scoring rule for assessing the quality of the predictions is the
SPE in Table~\ref{table:results-low-pass-SPE} and the IS at level~95\%
in Table~\ref{table:results-low-pass-IS}. %
(A similar table, not shown here, is presented in the SM for the
CRPS.) %

For comparison, Table~\ref{table:results-low-pass-SPE} also provides
the optimal values $R^{\star}$ obtained by direct minimization of the
score~\eqref{eq:gof}. %
They can be used to assess the loss of predictive accuracy of the
selected models, which are constructed using a limited number of
observations, with respect to the best model that could have been
obtained if the test data had also been used to select the parameters.

Table~\ref{table:results-low-pass-SPE} and
Table~\ref{table:results-low-pass-IS} support the fact that, for this
particular problem, the NLL and NLL/SPE criteria may be seen as
slightly better choices for selecting $\theta$ in terms of the SPE and
the IS scores, %
both for a prescribed regularity~$\nu$ and when $\nu$~is selected
automatically (the NLL/SPE being only available for the latter
case). %
The other selection criteria, however are
never \emph{very} far behind---at most a factor of approximately two,
for both metrics. %
Elements provided as SM show similar findings using the CRPS
validation score. %

Strikingly enough, for both scoring rules, the variations of the
average score are much larger when the model changes (with $\nu$) than
when the selection criterion changes. %
If a Matérn covariance function with fixed regularity is used, as is
often done in practice, then the best results are obtained for all
criteria when $\nu = 5/2$. %
The values of~$R^*$ (Table~\ref{table:results-low-pass-SPE}) confirm
that these are indeed the best fixed-$\nu$ models on this problem for
the SPE score. %
Since these optimal values were not known beforehand, it is a relief
to see (cf. last row of each table) that comparable performances can
be achieved on this problem by selecting~$\nu$ automatically.

Finally, we can also study the impact of the selection criterion and
that of $\nu$ on confidence intervals at level 95\%. This is a useful
complement to the results obtained using the IS for model
validation. For a Gaussian predictive distribution $P_{\theta, x}$ at
$x\in\XX$ with mean $\mu_{\theta}(x)$ and variance
$\sigma_{\theta}^2(x)$, the natural confidence interval at level 95\%
is
\begin{equation}
  I_{\theta}^{0.95}(x) = \left[ \mu_{\theta}(x)-1.96\sigma_{\theta}(x),\,
    \mu_{\theta}(x)+1.96\sigma_{\theta}(x) \right]\,.
\end{equation}
Then, denoting by $z:\XX\to\RR$ the unknown data generating function,
and given a test set
\begin{equation}
  \left\{ x_i^{\rm test}, i=1\,\ldots,\,N \right\}\in\XX^{N}
\end{equation}
at which the values of $z$ are known, define the \emph{empirical
  coverage} as
\begin{equation}
  C_{0.95}(\theta) = \frac{1}{N}\sum_{i=1}^N \one_{ z(x_i^{\rm test}) \in
    I_{\theta}^{0.95}(x_i^{\rm test})}\,.
\end{equation}
Table~\ref{tab:coverage_s2pi} reports the averaged values of
$C_{0.95}(\theta)$ over the $M$ repetitions for the data generating
function~$s_{2\pi}$.  Notice that when $\nu$ is large, empirical
coverages are low, indicating that the confidence intervals are
probably too small. Notice also that the NLL selection criterion tends
to produce well calibrated confidence intervals on this test function
for $\nu = 5/2$ or when $\nu$ is selected automatically.

\begin{table}[htbp]
  \caption{%
    Averages over the repetitions of the empirical coverage
    $C_{0.95}(\theta)$ for the function~$s_{2\pi}$.  Using
    $J_n^{\mathrm{NLL/SPE}}$ gives $C_{0.95}(\theta) = 0.93$.
  }
  \begin{center}
    \myTabConfig
    \begin{tabular}{| l | c | c | c | c | c | r |}
      \hline
      $C_{0.95}$
      & NLL
      & LOO-SPE
      & LOO-NLPD
      & LOO-CRPS
      & GCV\\
      \hline
      $\nu = 1/2$
      & \cellcolor[gray]{0.8412958265720641} $0.993$
      & \cellcolor[gray]{0.9650612011008065} $0.923$
      & \cellcolor[gray]{0.9719973425505999} $0.929$
      & \cellcolor[gray]{0.93688464244293}   $0.890$
      & \cellcolor[gray]{0.9758010854353633} $0.932$ \\
      \hline
      $\nu = 3/2$
      & \cellcolor[gray]{0.8573771384738481} $0.992$
      & \cellcolor[gray]{0.9551213550363553} $0.912$
      & \cellcolor[gray]{0.9716306809004309} $0.929$
      & \cellcolor[gray]{0.9694312564458032} $0.927$
      & \cellcolor[gray]{0.9799386762086273} $0.936$ \\
      \hline
      $\nu = 5/2$
      & \cellcolor[gray]{0.972139220238924}  $0.965$
      & \cellcolor[gray]{0.9320829105620877} $0.883$
      & \cellcolor[gray]{0.9427614866103643} $0.898$
      & \cellcolor[gray]{0.9393532085892835} $0.893$
      & \cellcolor[gray]{0.9484004690136385} $0.905$ \\
      \hline
      $\nu = 7/2$
      & \cellcolor[gray]{0.9694510312330505} $0.927$
      & \cellcolor[gray]{0.9079807302203421} $0.842$
      & \cellcolor[gray]{0.9058032463640817} $0.838$
      & \cellcolor[gray]{0.8987794265421399} $0.823$
      & \cellcolor[gray]{0.9054435573314525} $0.837$ \\
      \hline
      $\nu = 9/2$
      & \cellcolor[gray]{0.9292347680568753} $0.879$
      & \cellcolor[gray]{0.8899126732865678} $0.802$
      & \cellcolor[gray]{0.8852070362348582} $0.790$
      & \cellcolor[gray]{0.8773766792061743} $0.768$
      & \cellcolor[gray]{0.8769133136243806} $0.767$ \\
      \hline
      $\nu = 13/2$
      & \cellcolor[gray]{0.901321065937176}  $0.828$
      & \cellcolor[gray]{0.8751548283023809} $0.762$
      & \cellcolor[gray]{0.8684227429474504} $0.741$
      & \cellcolor[gray]{0.8622022456449236} $0.720$
      & \cellcolor[gray]{0.8644266923096718} $0.728$ \\
      \hline
      $\nu = 25/2$
      & \cellcolor[gray]{0.8843180958452838} $0.788$
      & \cellcolor[gray]{0.8715892075849854} $0.751$
      & \cellcolor[gray]{0.8604211719747145} $0.714$
      & \cellcolor[gray]{0.856669437611695}  $0.700$
      & \cellcolor[gray]{0.8590589267892087} $0.709$ \\
      \hline
      $\nu = \infty$
      & \cellcolor[gray]{0.8753623256525705} $0.763$
      & \cellcolor[gray]{0.8697529707554584} $0.745$
      & \cellcolor[gray]{0.8607212045481611} $0.715$
      & \cellcolor[gray]{0.8556462332582171} $0.696$
      & \cellcolor[gray]{0.8560871813480359} $0.698$ \\
      \hline
      $\nu \in \{ 1/2 , \cdots,\infty \}$
      & \cellcolor[gray]{0.9764513954518466} $0.963$
      & \cellcolor[gray]{0.899381675271142}  $0.824$
      & \cellcolor[gray]{0.9057477279837833} $0.838$
      & \cellcolor[gray]{0.8995254984166181} $0.824$
      & \cellcolor[gray]{0.9010795026154219} $0.828$ \\
      \hline
    \end{tabular}
  \end{center}
  \label{tab:coverage_s2pi}
\end{table}

\subsubsection{Statistical analysis of the benchmark results}
\label{sec:stat_bench}

Tables similar to Tables~\ref{table:results-low-pass-SPE}
and~\ref{table:results-low-pass-IS} have been produced %
for all the $(4 + 12) \times 3 = 48$ combinations of problem and design size, %
for the three scoring rules (SPE, CRPS and IS). %
We present in this section some graphical summaries and statistical
analyses of these results. %
The individual tables are provided in the~SM.

\paragraph{Comparison of the selection criteria} %
Figure~\ref{fig:calibrated_criteria} compares the distributions of the
average values of the SPE and IS scores for all selection criteria, in the case of a Matérn covariance
function with automatically selected regularity. %
As a preliminary observation, note that for most criteria (except GCV) the variations
of the ratio $R / R_0$ for the SPE score remain under four (second horizontal dashed
line), %
which is to be compared to the possibly large variations due to a poor choice of
covariance models (see the right plot in Figure~\ref{fig:model_boxplots} below).

\setlength{\overfullrule}{0pt}
\begin{figure}
  \hspace{-0.2cm}
  \begin{minipage}[b]{0.5\textwidth}
    \scalebox{0.61}{\input{\grphPATH calibrated_criteria_spe.pgf}}
    \label{fig:criteria_spe}
  \end{minipage}
  \begin{minipage}[b]{0.5\textwidth}
    \scalebox{0.61}{\input{\grphPATH calibrated_criteria_is.pgf}}
    \label{fig:criteria_is}
  \end{minipage}
  \caption{%
    Box plots of $R/R_0$ for different selection criteria, in the case where $\nu$~is automatically selected. %
    Each box plot is constructed using all problems and design sizes ($16 \times 3 = 48$ combinations).
      For a given problem, $R$ denotes the score, averaged over the repetitions, %
    and $R_0$ stands for the best value of~$R$ (among
    all models and selection criteria). %
    The left (resp. right) panel uses the $S^\SPE$
    (resp. $S_{0.95}^{\mathrm{IS}}$) as quality assessment
    criterion. %
    The box plots are sorted according to their upper whisker. %
    Grey dashed lines: $R / R_0 = 2, 4, 6, 8, 10$.
  }
  \label{fig:calibrated_criteria}
\end{figure}
\setlength{\overfullrule}{5pt}

A closer look at Figure~\ref{fig:calibrated_criteria} reveals that the
rankings of criteria obtained for both scoring rules are almost
identical. %
The ranking for the CRPS scoring rule (not shown) is the same as the
one for SPE. %
GCV provides the worst performance for all scoring rules, followed by
LOO-NLPD, while NLL dominates the ranking (for all scoring rules as
well).

\begin{remark}
  Observe in Figure~\ref{fig:calibrated_criteria} that LOO-SPE is,
  surprisingly, much less accurate than NLL according
  to~$S^{\SPE}$. %
  More generally, choosing a scoring rule~$S$ for the LOO criterion
  does not guarantee the highest precision according to this
  particular score.
\end{remark}

\paragraph{Comparison of the covariance models} %
Figure~\ref{fig:model_boxplots} compares the average values
of~$R(\theta; S^\SPE)$ when the NLL selection criterion is used on %
the set of GKLS problems, which have low regularities, and on the set
of low-pass filter problems, which contains very smooth instances. %

\setlength{\overfullrule}{0pt}
\begin{figure}
  \hspace{-3mm}
  \begin{minipage}[b]{0.5\textwidth}
    \scalebox{0.78}{\input{\grphPATH ML_MSPE_toms_5d_p.pgf}}
    \label{fig:toms_p_5d}
  \end{minipage}
  \begin{minipage}[b]{0.5\textwidth}
    \scalebox{0.78}{\input{\grphPATH ML_MSPE_lowpass_p.pgf}}
    \label{fig:lowpass_p}
  \end{minipage}
  \caption{%
    Box plots of $R/R_0$ using $\JNLL_n$ as selection criterion and
    $S^\SPE$ as quality assessment criterion, for different choices of
    regularity. %
    Left: all $5d$ GKLS problems. %
    Right: all low-pass filter problems. %
    Each box plot is constructed using all problems in the given class
    and all design sizes
    (left: $3 \times 3 = 9$ combinations; right: $4 \times 3 = 12$ combinations). %
    For a given problem, $R$ denotes the score, averaged over the
    repetitions, and %
    $R_0$ stands, as in Figure~\ref{fig:calibrated_criteria}, for the best value of $R$
    (among all models and selection criteria). %
    The box plots are sorted according to their upper whisker. %
    Grey dashed lines: $R / R_0 = 2, 4, 6, 8, 10$.}
  \label{fig:model_boxplots}
\end{figure}
\setlength{\overfullrule}{5pt}

Observe first that the fixed-$\nu$ models rank differently on these
two sets of problems, as expected considering the actual regularity of
the underlying functions: low values of~$\nu$ perform better on the
GKLS problems and worse on the low-pass filter case. %
Furthermore, it appears that underestimating the regularity (on the
low-pass filter case) has much more severe consequences than
overestimating it (on the GKLS problems) according to the SPE score,
and as suggested by the theoretical results of
\cite{stein1999:_interpolation_of_spatial_data},
\cite{narcowich_et_al2006:_evasion_theorems}, \cite{tuo2020:_kriging}
and~\cite[Section 6]{scheuerer_et_al2013:_interpolation_det_stoch}.

Another important conclusion from Figure~\ref{fig:model_boxplots} is
that very good results can be obtained by selecting the regularity
parameter~$\nu$ automatically, jointly with the other parameters
(using the NLL criterion in this case). %
On the GKLS problems, the results with selected~$\nu$ are not far from
those of the best fixed-$\nu$ model under consideration ($\nu =
3/2$); %
in the low-pass filter case, they are even better than those obtained
with the best fixed-$\nu$ models ($\nu = 5/2$ or~$7/2$). %
In other words, the regularity needs not be known in advance to
achieve good performances, which is a very welcome practical result. %
This conclusion is also supported, for NLL, by the additional results
provided in the SM for the other problems and for the three scoring
rules.

Concerning the other selection criteria the situation is more
contrasted (see SM): %
the automatic selection of~$\nu$ using these criteria still performs
very well for smooth problems, but not always, in particular with GCV,
for the less regular problems of the GKLS class. %
This is especially true when the sample size is small ($n = 10d$).

\paragraph{Robustness}
LOO-SPE is commonly claimed in the literature (see, notably,
\cite{bachoc:2013:cv-ml-misspec}) to provide a certain degree of
robustness with respect to model misspecification. %
According to this claim, LOO-SPE would be expected to somehow mitigate
the loss of predictive accuracy with respect to likelihood-based
approaches incurred by an ill-advised choice of covariance function. %
Our detailed results (see~SM) suggest that this effect indeed exists
when the regularity is severely under-estimated (e.g., $\nu = 1/2$ for
the low-pass filter problems), but is actually quite small, and should
not be used to motivate the practice of setting $\nu$ to an arbitrary
value. %
A similar effect exists for LOO-CRPS, LOO-NLPD and GCV as well. %
Quite surprisingly, NLL turns out to be more robust than LOO-SPE (and
the other criteria) in the case of oversmoothing.

\paragraph{Coverage}

To further complement the study, we analyze the results in terms of
empirical coverage, as defined in Section~\ref{sec:closer-look}.
Figure~\ref{fig:coverage} shows the variations of the empirical
coverage as a function of the covariance model on the GKLS test
functions.  As expected, choosing large values of $\nu$ for these test
functions tends to produce undercoverage, that is, (too) small
confidence intervals, or slightly overconfident predictions.
Figure~\ref{fig:coverage} also shows the effect of the selection
criterion (when $\nu$ is automatically selected) on the whole set of
test functions. As in Section~\ref{sec:closer-look}, it seems that the
NLL selection criterion yields better calibrated confidence intervals.

\setlength{\overfullrule}{0pt}
\begin{figure}
  \hspace{-3mm}
  \begin{minipage}[b]{0.5\textwidth}
    \scalebox{0.62}{\input{\grphPATH ML_COVERAGE_toms_5d_p.pgf}}
  \end{minipage}
  \begin{minipage}[b]{0.5\textwidth}
    \scalebox{0.62}{\input{\grphPATH calibrated_criteria_coverage.pgf}}
  \end{minipage}
  \caption{%
    Box plots of $C_{0.95} \left(\theta\right)$ (averaged over the
    repetitions for each test function).  Left: for several
    regularities and using the NLL criterion on all $5d$ GKLS
    problems; right: for several selection criteria and for $\nu$
    automatically selected, on the whole set of test functions.}
  \label{fig:coverage}
\end{figure}
\setlength{\overfullrule}{5pt}

\paragraph{Sensitivity analysis} %
We observed in Section~\ref{sec:closer-look} and in paragraphs above
that the choice of the model---more specifically, of the regularity
parameter of the Matérn covariance function---was more important than
that of a particular selection criterion. %
To confirm this finding, a global sensitivity analysis of the
logarithm of the score has been performed for each combination
of problem and design size. %
The score, for a given scoring rule, depends on three discrete
factors: the selection criterion, the regularity~$\nu$ of the Matérn
covariance function, and the %
repetition index (recall that for each problem we have a collection of
$M$~repetitions, with $M = 100$ or~$15$ depending on the problem). %
Only fixed values of~$\nu$ are considered in this analysis (no automatic
selection), and therefore the NLL/SPE method, which coincides with NLL
for fixed~$\nu$, is not included as a level for the criterion factor. %
The factors are considered independent and uniformly distributed in the
definition of the Sobol' sensitivity indices.

We present in Figure~\ref{fig:sobol_spe}, for the case of the SPE scoring rule, the
first order Sobol' sensitivity indices with respect to~$\nu$ and
the total Sobol' sensitivity indices with respect to the selection
criterion, as functions of the total variance. %
Observe that, for the problems where the total variance is large, the
total index with respect to the selection criterion is
typically close to zero, which indicates a negligible impact of the
selection criterion on the variations of $R$. %
On the other hand, first order indices with respect to~$\nu$ are typically
above~$80\%$ on problems with large variations of~$R$. %
This indicates situations where the influence of~$\nu$ is very important.

Note also that it also means that the influence of $\nu$ dominates the
variability due to the randomness of the design over the repetitions,
which apparently counterbalances the conclusions of
\cite{chen_et_al2016:_ce_assess}, where large variations of the
prediction errors are reported when the design varies. %
However, \cite{chen_et_al2016:_ce_assess} is in the case of small
designs, and from a fully Bayesian point of view, the posterior
distribution of $\theta$ is very often very spread out in this case,
potentially leading to poor parameter selection, as also noted
in~\cite{benassi11:robust}. %
In our setting however, our primary focus is to study the problem of
model selection, and so we have chosen to consider ``informative''
designs in our experiments (using maximin Latin hypercube sampling and
design sizes $n$ with $n \geq 10d$).

Similar conclusions hold for the other scoring rules (results not
shown, see~SM).

\begin{figure}
  \begin{center}
    \resizebox{\textwidth}{!}{\input{\grphPATH sobol_nu_spe.pgf} }
    \resizebox{\textwidth}{!}{\input{\grphPATH sobol_crit_spe.pgf} }
  \end{center}
  \caption{%
    Sensitivity analysis of $\log_{10}(R)$ for $S^\SPE$, for each problem and design size.  The
    top plots show the first-order Sobol' indices for~$\nu$, i.e., the
    contribution of $\nu$ to the variance of $S^\SPE$ measured as the
    ratio of the variance of the expectation of $S^\SPE$ when $\nu$ is
    fixed to the total variance of $S^\SPE$. %
    Values close to one tells us that the variations of $\nu$ explain
    almost all the variations of $S^\SPE$. %
    The bottom plots show the total Sobol' indices of the selection
    criterion, i.e., the variance explained by the selection
    criterion, taking every interactions into account. %
    Values close to zero tells us that the variations of $S^\SPE$ are
    not explained by variations of the selection criterion.  Each
    point represents the variations of $\log_{10}(R)$ for one of the
    $16$ problems from Section~\ref{sec:test-functions-from}, split by
    design size. %
    The model explains almost all the variations for problems
    exhibiting significant fluctuations of $\log_{10}(R)$ (at the
    right of the figure). %
  }
\label{fig:sobol_spe}
\end{figure}

\section{Conclusions}
\label{sec:conc}

A large variety of selection criteria for Gaussian process models is
available from the literature, with little theoretical or empirical
guidance on how to choose the right one for applications. %
Our benchmark study with the Matérn family of covariance functions in
the noiseless (interpolation) case %
indicates that the NLL selection criterion---in other words, the ML
method---provides performances that are, in most situations and for
all the scoring rules that were considered (SPE, CRPS and IS at~95\%),
better than or comparable to those of the other criteria. %
Considering that all the criteria tested in the study have
a similar computational complexity, this provides a strong empirical
support to the ML method---which is already the \emph{de facto}
standard in most statistical software packages implementing Gaussian
process interpolation.

Another important lesson learned from our benchmark study is that the
choice of the family of models, and in particular of the family of
covariance functions, has very often a bigger impact on performance
than that of the selection criterion itself. %
This is especially striking when the actual function is smooth, and
very irregular covariance function such as the Matérn covariance with
regularity~$1/2$ is used to perform Gaussian process interpolation. %
In such a situation, NLL is actually outperformed by other criteria
such as LOO-SPE or LOO-CRPS, which thus appear to be more ``robust to
model misspecification''. %
However, the small gain of performance, which is achieved by using
LOO-SPE or LOO-CRPS instead of NLL in this case, is generally
negligible with respect to the much larger loss induced by choosing an
inappropriate covariance function in the first place.

Our final recommendation, supported by the results of the benchmark,
and which is in line with the recommendation
of~\cite{stein1999:_interpolation_of_spatial_data}, is therefore to
select the regularity of the covariance function automatically,
jointly with the other parameters, using the NLL criterion. %
A minimal list of candidate values for the regularity parameter should
typically include~$1/2$, $3/2$, $5/2$, $7/2$ and~$+\infty$ (the
Gaussian covariance function). %
Should a situation arise where a default value of~$\nu$ is
nevertheless needed, our recommendation would be to choose a
reasonably large value such as~$\nu = 5/2$ or~$\nu = 7/2$, %
since underestimating~$\nu$ has been seen in
Section~\ref{sec:stat_bench} to have much more severe consequences
than overestimating it a little. %
Nevertheless, we stress that overestimating~$\nu$ is not only
sub-optimal, but it is also likely to produce confidence intervals
with undercoverage.

More generally, our numerical results support the fact that choosing
carefully a suitable family of Gaussian process models
is extremely important: %
poor decisions at this stage are indeed likely to induce large losses
in performance that no choice of selection criterion can
counterbalance. %
We have focused in this article on the case of a constant mean with an
anisotropic Matérn covariance function, with fixed regularity or
not. %
We believe---but it should be confirmed by further numerical experiments---that this conclusion would not be altered significantly if
other families of Gaussian process models with a comparable number of
parameters (e.g., affine mean fuction, tensor-product covariance functions, or the
compactly supported covariance functions proposed by
\cite{wendland95:_piecew}) were considered instead, or if other
noise-free settings (multivariate outputs, multi-fidelity\ldots) were
addressed.

However, it should be kept in mind that the study focuses on cases
where the number of parameters is small with respect to the number of
observations %
(in particular, we considered GPs with an anisotropic stationary
Matérn covariance function, which have $d+2$ parameters, and we took
care of having $n \gg d$). %
When $n$ is small with respect to $d$, or when the number of
parameters increases, it seems to us that a fully Bayesian approach
could help, or that other selection criteria should be considered,
with the introduction of regularization terms, for instance.

For future work, it would be of course very interesting to consider
the performance of using selection criteria against a fully Bayesian
approach. %
Another direction would be to extend this study to the case of
regression, which is also used in many applications, when dealing with
stochastic simulators, for instance.

\section*{Acknowledgments}
The authors thank Mona Fuhrländer and Sebastian Schöps for kindly
providing the data for the ``Design of a low-pass filter'' example. %
This work was funded by the French %
\textit{Agence Nationale de la Recherche et de la Technologie} (ANRT)
under a CIFRE grant.

\bibliographystyle{siamplain}
\bibliography{spetit-gpparam-biblio}

\end{document}